\documentstyle[12pt]{article}
\input epsf

\newcommand{\resection}[1]{\setcounter{equation}{0}\section{#1}}
\newcommand{\appsection}[1]{\setcounter{equation}{0}\section*{Appendix}}

\def\d{\delta}
\def\t{\theta}

\def\l{\lambda}

\def\e{\epsilon}

\def\s{\sigma}
\def\SM{$S$-matrix}
\def\p{\pi}

\def\Gam{\Gamma}

\def\be{\begin{equation}}
\def\ee{\end{equation}}
\def\bea{\begin{eqnarray}}
\def\eea{\end{eqnarray}}
\def\beano{\begin{eqnarray*}}
\def\eeano{\end{eqnarray*}}
\def\bd{\begin{displaystyle}}
\def\ed{\end{displaystyle}}
\def\ba{\begin{array}}
\def\ea{\end{array}}
\def\nls{nl$\sigma$ }

\def\lab{\label}
\def\nn{\nonumber}
\def\nv{\vec {\bf n}}

\def\pd{\partial}

\def\Jrnl#1#2#3#4{{#1} {\bf #2}, (#4) #3}

\def\NPB{{ Nucl. Phys.} \bf B}
\def\PLB{{ Phys. Lett.}  \bf B}

\def\PRD{{ Phys. Rev.} \bf D}

\def\IJMPA{{ Int. Jour. Mod. Phys.} \bf A}

\def\AnPh{Ann. Phys. }

\begin{document}
\oddsidemargin 5mm
\setcounter{page}{0}
\newpage     
\setcounter{page}{0}
\begin{titlepage}
\begin{flushright}
LPENSL-TH-12/98\\
IC/98/223
\end{flushright}
\vspace{0.5cm}
\begin{center}
{\large {\bf Boundary $S$-matrix for the Gross-Neveu Model}}
\footnote{Work done under partial support of the 
EC TMR Programme {\noindent \em Integrability, non--perturbative effects 
and symmetry in Quantum Field Theories}, grant FMRX-CT96-0012, and of 
CNRS (France).\\ \indent \ {}* URA 1325 du CNRS, associ\'ee \`a l'Ecole 
Normale Sup\'erieure de Lyon.}\\
\vspace{1.5cm}
{\bf A. De Martino}\footnote{\noindent \tt{email:ademarti@ens-lyon.fr}}\\
{\em Laboratoire de Physique*, Groupe de Physique Th\'eorique}\\
{\em ENS Lyon, 46 all\'ee d'Italie, 69364 Lyon Cedex 07, France}\\
\vspace{0.8cm}
{\bf M. Moriconi}\footnote{\noindent \tt{email:moriconi@ictp.trieste.it}}\\
{\em The Abdus Salam International Centre for Theoretical Physics}\\
{\em Strada Costiera 11, 34014 Trieste, Italy}\\
{\em and}\\
{\em Istituto Nazionale di Fisica Nucleare, Sezione di Trieste}\\
\end{center}
\vspace{6mm}
\begin{abstract}
\noindent
We study the scattering theory for the Gross-Neveu model on the half-line.
We find the reflection matrices for the elementary fermions, and by fusion 
we compute the ones for the 
two-particle bound-states, showing that they satisfy non-trivial bootstrap 
consistency conditions. 
We also compute more general reflection matrices for the Gross-Neveu model 
and the nonlinear sigma model, and argue that they correspond to the 
integrable boundary conditions we identified in our previous paper 
\cite{MorDeMar}. 
\vspace{3cm}
\end{abstract}
\vspace{5mm}
\end{titlepage}

\newpage
\setcounter{footnote}{0}
\renewcommand{\thefootnote}{\arabic{footnote}}

\resection{Introduction}

After the seminal paper by Ghoshal and Zamolodchikov \cite{gzam} (see
also \cite{FrKob}) on integrable quantum field theories with 
boundary, a lot of work has been done extending it, 
especially in the analysis of different models. 
In this paper we study one of these extensions, namely the $O(N)$
Gross-Neveu (GN) model \cite{GrNev} on the half-line, which is closely 
related to the boundary $O(N)$ non-linear sigma (nl$\s$) model studied by 
Ghoshal in \cite{Ghoshal}.
Recently \cite{MorDeMar} we have found new integrable boundary conditions 
(bc's) for the GN and \nls models, based on the microscopic (lagrangian) 
description
of these models. Here we find general (diagonal) solutions for the boundary 
Yang-Baxter equation and propose a one-to-one correspondence between these 
solutions and the boundary conditions we found in \cite{MorDeMar}.

This paper is organized as follows. In the next section we briefly
review the GN and \nls models, and their exact $S$-matrices;   
in section 3 we find the exact reflection matrices for the GN model with
the physically interesting free (Neumann) and fixed boundary conditions;
in section 4 we solve the boundary Yang-Baxter equation for the more
general boundary conditions we found in our previous paper
\cite{MorDeMar}, and establish the correspondence between those 
boundary conditions and the 
reflection matrices we find;
in the final section we present our conclusions.
We also have an appendix where we write the exact amplitudes
for the scattering between bound-states of elementary fermions in the GN
model \cite{ZamZam,karthun}. Their computation is a simple exercise in 
fusion, but since we need their explicit form in this paper, we give 
the results in the appendix.

\resection{The Gross-Neveu Model}

The GN model is defined by the following lagrangian
\be
{\cal{L}}_{gn}=\bar{\psi} 
i \partial\!\!\!\slash
\psi+\frac{g^2}{2}(\bar{\psi}\psi)^2
\ , \ \label{gn}
\ee
where $\psi$ is a $N$-component massless Majorana fermion in the
fundamental representation of $O(N)$. 
In the above equation $\bar{\psi}\psi$ should be understood as 
$\sum_{i=1}^N\bar{\psi}^{i}\psi^{i}$ and so on. 
In this paper we will refer to the $O(N)$ GN model simply as GN model. 
It is useful to write the GN model lagrangian in light-cone coordinates, 
where it reads
\be
{\cal L}_{gn}=2 \psi_+^i i\pd_- \psi_+^i +2 \psi_-^i i\pd_+ \psi_-^i +
2 g^2(\psi_+^i\psi_-^i)^2 \ . \
\ee

The particle spectrum of the GN model \cite{dhn} is composed by the
$O(N)$ vector multiplet of elementary particles included in the
lagrangian (the ``elementary fermions'') and a set of $O(N)$
multiplets (scalar and higher rank antisymmetric tensors) of
increasing mass, which can be thought of as bound-states of a number
of elementary fermions\footnote{In fact, the spectrum contains also
kink states, associated to the spontaneous breaking of the chiral
symmetry \cite{Witten}.}. 
In what follows we will restrict our considerations to the
sector of the theory containing only the elementary fermions (denoted
by the symbols $A_i$, $i=1,\ldots,N$) and the two-fermion bound-states: the
isoscalar particle (denoted by $B$), corresponding to the bound-state
in the isoscalar channel of the \SM of elementary fermions,
and the antisymmetric multiplet 
(denoted by $B_{ij}$), corresponding to the bound-state in the
antisymmetric channel.

The exact \SM for the elementary fermions was found by Zamolodchikov and
Zamolodchikov in \cite{ZamZam} and we quote it here for further
reference. The Faddeev-Zamolod\-chikov algebra is
\bea
&&A_i(\t_1)A_j(\t_2)=\delta_{ij}\s_1(\t_{12})\sum_{k=1}^n A_k(\t_2)A_k(\t_1) + 
\nn \\
&&\phantom{A_i(\t_1)A_j(\t_2)=}
+\s_2(\t_{12})A_j(\t_2)A_i(\t_1)+
\s_3(\t_{12})A_i(\t_2)A_j(\t_1)\, ,
\label{FZA}
\eea
where $\t_{12}=\t_1-\t_2$\,. The  $\s_i(\t)$ are given by
\be
\s_1(\t)=-\frac{i\l}{i \p -\t}\s_2(\t) \qquad \ , \ \qquad
\s_3(\t)=-\frac{i\l}{\t}\s_2(\t)\, ,
\label{GNampl}
\ee
and
\be
\s_2(\t)=\frac{\sinh\t+i\sin\l}{\sinh\t-i\sin\l}\,\s^0_2(\t)\, ,
\label{s2}
\ee
with
\be
\s^{0}_2(\t)=
\frac{\Gam(\frac{\l}{2\p}-\frac{i\t}{2\p})\,\Gam(\frac{1}{2}-\frac{i\t}{2\p})}
{\Gam(\frac{1}{2}+\frac{\l}{2\p}-\frac{i\t}{2\p})\,\Gam(-\frac{i\t}{2\p})}
\frac{\Gam(\frac{1}{2}+\frac{\l}{2\p}+\frac{i\t}{2\p})\,\Gam(1+\frac{i\t}{2\p})}
{\Gam(1+\frac{\l}{2\p}+\frac{i\t}{2\p})\,\Gam(\frac{1}{2}+\frac{i\t}{2\p})}\
, \
\label{s2gam}\\
\ee
where we introduced $\l=2\p/(N-2)$. The 
pole at $\t=i\l$ in the CDD prefactor in equation 
\ref{s2} gives rise, by bootstrap, to the full exact spectrum of bound-states. 

The \SM elements for $B$ and $B_{ij}$ particles can be obtained from the 
elementary fermion \SM by fusion, using the identity
\cite{ZamZam}
\be
A_i(\t-i\l/2)\,A_j(\t+i\l/2)=\d_{ij}\,B(\t)+\sqrt{(N-4)}\,B_{ij}(\t) \ . \ 
\label{fusion}
\ee
A partial list of these amplitudes can be found in the appendix.

The GN model is closely related to the $O(N)$ \nls model; 
they share very similar properties, and in particular the 
elementary particles in both models have the same $S$-matrix, up 
to a CDD factor.
The $O(N)$ \nls model is defined by the following lagrangian
\be
{\cal L}_{nl\s}=\frac{1}{2 g_0} \pd_\mu \nv \cdot \pd^\mu \nv
\ , \ \label{nlsm}
\ee
where $\nv$ is a vector in $N$-dimensional space, subject to the constraint
$\nv \cdot \nv=1$. The exact \SM for the elementary particles in the
\nls model is given by equations \ref{FZA}-\ref{s2gam}
with $\s_2$ substituted by $\s_2^0$ \cite{ZZAnn}, and consequently there are 
no bound-states in this model.

\resection{Reflection Matrices}

In this section we will compute the reflection matrices for the GN model with
free and fixed boundary conditions, following Ghoshal's analysis for the 
\nls model \cite{Ghoshal}. Before we proceed let us note that what Ghoshal
means by fixed boundary condition is not what we meant by Dirichlet boundary
condition in our previous paper \cite{MorDeMar} (and in this one). 
Ghoshal's condition corresponds to leaving
only {\em one} component of the field $\nv$ fixed at the boundary, while
by Dirichlet boundary condition it should be understood that the field $\nv$ 
is fixed at the boundary, $\nv(x,t)|_{x=0}=\nv_0$.

Since the \SM of the GN model is, up to a CDD factor, the same as
the \SM of the $O(N)$ \nls model, 
the boundary Yang-Baxter equation (BYBE) 
for the two models is exactly the same,
and so the reflection matrix of the GN model is given by the one of the 
nl$\s$ model multiplied by the appropriate CDD 
factors\footnote{This will give us a minimal solution, without any 
extra poles in the physical strip.}. 
This indicates that there is a one-to-one correspondence between the integrable
boundary conditions for the GN and \nls model. Similarly to the bulk case, 
the different physics of these two models resides in these CDD prefactors. 

Let us summarize the results of the analysis of integrable boundary conditions
for the GN model \cite{MorDeMar}.
The action of the boundary GN model is given by
\be S_{bgn}= \int_{-\infty}^{0} dx
\int_{-\infty}^{\infty} dt \,\, {\cal L}_{gn} + \int_{-\infty}^{\infty} dt
\,\, {\cal L}_b \, , \ \label{bgn} 
\ee 
where ${\cal L}_b$ is the boundary action.
As we have shown in \cite{MorDeMar} the boundary lagrangian
\be {\cal L}_b=\sum_{i=1}^N
\frac{i}{2} \e_i \psi_+^i \psi_-^i \ , \ \label{Lb} 
\ee where
$\e_i=\pm 1$, preserves the integrability of the GN model at the quantum
level. The boundary condition derived from this action is
\be 
\psi_+^i|_{x=0}=\e_i \psi_-^i|_{x=0} \ . \ \label{bc} 
\ee 
Borrowing the terminology from the \nls model, we will refer to bc's with all 
$\e_i=+1$ or all $\e_i=-1$ respectively as Neumann and Dirichlet bc's.
Therefore, up to index reshuffling,
we have $N+1$ inequivalent boundary conditions, which have
to correspond to different solutions of the boundary Yang-Baxter
equation. Due to the fact that the boundary interaction does not
involve any flavor-changing terms, we should be able to find {\em
diagonal} solutions for the BYBE, which will be done in section 4.  In
this section, following Ghoshal analysis, we exhibit solutions for the free 
and fixed boundary conditions, which serves as a warm-up for the more general 
case.

\subsection{Free Boundary Condition}

As it follows from our discussion above, we should assume that the 
reflection amplitude for the elementary 
fermions is given by 
\be 
R_i^j(\t) \equiv R(\t)\, \delta_i^j \, .
\ee
Physically it means that we can not change the `flavor' of the 
fermion by scattering it off the boundary and that 
the amplitude of scattering does not depend on the index $i$. 
For the \nls model this ansatz corresponds to no interactions on the boundary 
\cite{Ghoshal}.
In the case of GN model, it corresponds both to the bc's in 
equation \ref{bc} with all $\e_i$ equal to ``$+$'' or all  $\e_i$ equal 
to ``$-$'', the difference between the two cases lying in a CDD factor.
Due to the similarity between GN and \nls model 
$S$-matrices, $R(\t)$ can be written as 
\be
R(\t) = f(\t)\, R_0(\t)\, ,
\lab{reflfree}
\ee
where $R_0(\t)$ is the reflection amplitude for the \nls model with free bc
computed by Ghoshal in  \cite{Ghoshal}:
\be
R_0(\t)=\frac{\Gamma(\frac{1}{2}+\frac{\l}{4\pi}-\frac{i\t}{2\pi})\,
\Gamma(1+\frac{i\t}{2\pi})\,
\Gamma(\frac{3}{4}+\frac{\l}{4\pi}+\frac{i\t}{2\pi})\,
\Gamma(\frac{1}{4}-\frac{i\t}{2\pi})}
{\Gamma(\frac{1}{2}+\frac{\l}{4\pi}+\frac{i\t}{2\pi})\,
\Gamma(1-\frac{i\t}{2\pi})\,
\Gamma(\frac{3}{4}+\frac{\l}{4\pi}-\frac{i\t}{2\pi})\,
\Gamma(\frac{1}{4}+\frac{i\t}{2\pi})}\, .
\lab{Gfc}
\ee
The prefactor $f(\t)$ is fixed by unitarity and boundary crossing-unitarity,
which generally read
\be
R_i^k(\t)\,R_k^j(-\t)=\delta_i^j \qquad \ , \ \qquad 
K^{ij}(\t)=S_{i'j'}^{ji}(2\t)\,K^{i'j'}(-\t)\, , 
\label{RConditions}
\ee
where $K^{ij}(\t)=C^{ii'}R^{j}_{i'}(\frac{i\pi}{2}-\t)$, and $C^{ij}$ is the
charge conjugation matrix. Equations \ref{RConditions} imply that $f(\t)$ 
should satisfy 
\bea
f(\t)f(-\t)&=&1\, , \lab{fdef1} \\
f(\frac{i \p}{2}-\t)&=&
\frac{\sinh 2\t+i\sin\l}{\sinh 2\t-i\sin\l}\,\, 
f(\frac{i \p}{2}+\t)\, .
\lab{fdef}
\eea
The minimal solution of \ref{fdef1}, \ref{fdef} can be found by elementary 
methods. In fact, there are two minimal solutions, with rather
different physical properties. They are given by 
\be
f(\t)=\Phi(\t)\,
\frac{\sinh\frac{1}{2}(\t+\frac{i\l}{2})\,
\sinh\frac{1}{2}(\t-\frac{i\l}{2}-\frac{i \p}{2})}
{\sinh\frac{1}{2}(\t-\frac{i\l}{2})\,
\sinh\frac{1}{2}(\t+\frac{i\l}{2}+\frac{i \p}{2})} \, , 
\lab{fdef2}
\ee
and
\be               
f(\t)=\Phi(\t)\,
\frac{\sinh\frac{1}{2}(\t+\frac{i\l}{2})\,
\sinh\frac{1}{2}(\t-\frac{i\l}{2}+\frac{i \p}{2})}
{\sinh\frac{1}{2}(\t-\frac{i\l}{2})\,
\sinh\frac{1}{2}(\t+\frac{i\l}{2}-\frac{i \p}{2})} \, ,
\lab{fdef3}
\ee
differing by the CDD factor
\be
F_{\mbox{\tiny CDD}}(\t)=
\frac{\tanh(\frac{i\pi}{4}-\frac{i\l}{4}+\frac{\t}{2})}
{\tanh(\frac{i\pi}{4}-\frac{i\l}{4}-\frac{\t}{2})}\, .
\ee
The first solution exhibits only a simple pole in the physical strip, at
$\t=i\l/2$, corresponding to the bound-state pole at $\t=i\l$ in the bulk
\SM. 
The second solution exhibits an additional simple pole at
$\t=i\pi/2-i\l/2$, meaning that the boundary state for this solution
contains a zero-rapidity single-particle contribution by the
particle $B$ \cite{gzam}.

In equations \ref{fdef2}, \ref{fdef3} $\Phi(\t)$ is a prefactor
satisfying
\be
\Phi(\t)\,\Phi(-\t)=1 \qquad \ , \ \qquad
\Phi(\frac{i \p}{2}-\t)=-\,\Phi(\frac{i \p}{2}+\t)\, .
\lab{Ising}
\ee
The above equations are exactly those for the reflection
matrix of the Ising model \cite{gzam}. Since we do not expect any free 
parameters in our reflection matrices (the boundary term in the lagrangian 
\ref{Lb} having no free parameters), we pick up the minimal solutions
corresponding precisely 
to the boundary conditions $\psi_+|_{x=0}=\pm \psi_-|_{x=0}$ for Ising 
fermions:
\be
\Phi_{+}=-i\coth\left(\frac{i\pi}{4}-\frac{\t}{2}\right)
\qquad \ , \ \qquad \Phi_{-}=
i\tanh\left(\frac{i\pi}{4}-\frac{\t}{2}\right)
\, .
\ee
Therefore, we propose that $R(\t)$ in equation \ref{reflfree} with $\Phi_+$ 
corresponds to Neumann bc and with $\Phi_-$ to Dirichlet bc.

As we noticed before, the reflection amplitude $R(\t)$ has a pole at 
$\t=\frac{i\l}{2}$, corresponding to the bound-state pole at $\t=i\l$ in 
the bulk \SM. 
Since the particles $B$ and $B_{ij}$ can be interpreted as bound-states 
of $A_iA_j$, we can use the boundary-bootstrap equation to compute the 
reflection amplitudes for them (see also \cite{MacKay}), assuming that the
boundary has no 
structure and consequently that the only non vanishing reflection 
factors are $R_B^B(\t)$ and $R_{B_{ij}}^{B_{ij}}(\t)$. 
Recall that in general, if the particle $A_b$ can be interpreted as a 
bound-state of $A_{a_1}A_{a_2}$ (corresponding to a pole in the bulk 
scattering amplitude at $\t=i u_{a_1a_2}^b$), the boundary \SM elements for 
$A_b$ can be obtained by taking the appropriate residue at the bound-state 
pole of the two-particle boundary \SM $R_{a_1a_2}^{a_1a_2}(\t_1,\t_2)$
\cite{gzam}:
\be
f_{a_1a_2}^b\, R_b^c(\t)=f_{c_1c_2}^c\, 
R_{a_2}^{b_2}(\t-i\bar u_{a_2 b}^{a_1})\,
S_{a_1b_2}^{b_1c_2}(2\t+i\bar u_{ba_1}^{a_2}-i\bar u_{a_2 b}^{a_1})\, 
R_{b_1}^{c_1}(\t+i\bar u_{ba_1}^{a_2})\, .
\label{bboot}
\ee
where $\bar u=\pi-u$ and $f_{a_1a_2}^b$ are the three-particle
on-shell
couplings defined by the residue of the bulk \SM at the bound state pole:
\be
S_{a_1a_2}^{c_1c_2}(\t)\simeq\frac{\t-iu_{a_1a_2}^b}
{i\,f_{a_1a_2}^b}f^{c_1c_2}_b\, .
\ee 

Notice that the fused reflection amplitude is 
manifestly unitary, and we need only check boundary crossing-unitarity.
A straightforward bootstrap computation gives
\bea
&&R_B^B(\t)= -\frac{(i\l+2\t)(i\pi+2\t)}{(2\t)(i\pi-2\t)} 
R(\t_-)\,R(\t_+) \s_2(2\t) \, ,\\
&&R_{B_{ij}}^{B_{ij}}(\t)= -\frac{i\l+2\t}{2\t}R(\t_-)\,R(\t_+))\s_2(2\t) \, ,
\eea
where $\t_{\pm}=\t\pm \frac{i\l}{2}$.

We can check the consistency of the bootstrap computation by verifying
that the appropriate boundary crossing-unitarity equation is satisfied
by these reflection amplitudes.
It can be written easily, but one should be careful
with factors coming from charge conjugation (see the appendix in \cite{DMM}). 
The final result is  
\be
K^{BB}(\t)=K^{BB}(-\t)S_{BB}^{BB}(2\t)+
\frac{N(N-1)}{2}K^{B_{ij}B_{ij}}(-\t) S_{B_{ij}B_{ij}}^{BB}(2\t)
\label{1bcu}
\ee
where $K^{BB}(\t)=R_B^B(\frac{i \p}{2}-\t)$ and
$K^{B_{ij}B_{ij}}(\t)=-2R_{B_{ij}}^{B_{ij}}(\frac{i \p}{2}-\t)$. By using the 
bulk amplitudes listed in the appendix, equation \ref{1bcu} can be easily 
shown to be satisfied.

\subsection{Fixed Boundary Condition}

Now let us consider the case where the first $N-1$ fermions satisfy the ``$+$''
bc and the $N$-th fermion satisfies the ``$-$'' bc.
In terms of reflection matrices, this situation is described by the ansatz
\bea
R_i^i(\t)&\equiv&R_1(\t)\, , \qquad i=1,\dots, N-1\, , \nn\\
R_N^N(\t)&\equiv& R_2(\t)\, ,
\label{fix}
\eea
which is the same as the one for fixed bc in the \nls model 
considered by Ghoshal, and therefore we follow his analysis closely. 
The amplitudes \ref{fix} can be written as
\bea
&&R_1(\t) = f(\t)\,R_1^0(\t)\, , \nn \\
&&R_2(\t) = f(\t)\,R_2^0(\t)\, ,
\eea
where $f(\t)$ is given by \ref{fdef2} and $R_1^0$ and $R_2^0$ are the 
amplitudes for the \nls model. From the BYBE Ghoshal found
\be
X(\t)\equiv\frac{R^0_1(\t)}{R^0_2(\t)}=\frac{i \p -2\t}{i \p+2\t} \,\, .
\label{condition}
\ee
and by solving unitarity and crossing-unitarity 
\bea
&&R_1^0(\t)=-\frac{\Gamma(\frac{1}{2}+\frac{\l}{4\pi}-\frac{i\t}{2\pi})
\Gamma(1+\frac{i\t}{2\pi})\,
\Gamma(\frac{1}{4}+\frac{\l}{4\pi}+\frac{i\t}{2\pi})\,
\Gamma(\frac{3}{4}-\frac{i\t}{2\pi})}
{\Gamma(\frac{1}{2}+\frac{\l}{4\pi}+\frac{i\t}{2\pi})\,
\Gamma(1-\frac{i\t}{2\pi})\,
\Gamma(\frac{1}{4}+\frac{\l}{4\pi}-\frac{i\t}{2\pi})\,
\Gamma(\frac{3}{4}+\frac{i\t}{2\pi})}\, , \nn \\
&&R_2^0(\t)=-\frac{\Gamma(\frac{1}{2}+\frac{\l}{4\pi}-\frac{i\t}{2\pi})\,
\Gamma(1+\frac{i\t}{2\pi})\,
\Gamma(\frac{1}{4}+\frac{\l}{4\pi}+\frac{i\t}{2\pi})\,
\Gamma(-\frac{1}{4}-\frac{i\t}{2\pi})}
{\Gamma(\frac{1}{2}+\frac{\l}{4\pi}+\frac{i\t}{2\pi})\,
\Gamma(1-\frac{i\t}{2\pi})\,
\Gamma(\frac{1}{4}+\frac{\l}{4\pi}-\frac{i\t}{2\pi})\,
\Gamma(-\frac{1}{4}+\frac{i\t}{2\pi})}\, . \nn
\eea
Using these reflection amplitudes we can compute, as before, the reflection 
factors for the two-particle bound-states. Notice that in this case 
we can obtain $B$ by fusing fermions satisfying ``$+$'' bc:
\be
R_B^B(\t)=R_1(\t_-)\,R_1(\t_+)
\left[(N-1)\,\s_1(2\t)+\s_2(2\t)+\s_3(2\t)\right]+
R_1(\t_-)\,R_2(\t_+)\s_1(2\t) \ ; \ \label{rbb1}
\ee
but we can also write $B$ as the fusion of the $N$-th fermion, 
which satisfies ``$-$'' bc:
\be
R_B^B(\t)=R_2(\t_-)\,R_2(\t_+)
\left[\s_1(2\t)+\s_2(2\t)+\s_3(2\t)\right]+
R_2(\t_-)\,R_1(\t_+)\,(N-1)\,\s_1(2\t) \ . \ \label{rbb2}
\ee
These two expressions for $R_B^B(\t)$ have to be equal, and this provides a
non-trivial consistency condition for the boundary-bootstrap:
\bea
&&X(\t_-)\,X(\t_+)\left[ (N-1)\,\s_1(2\t)+\s_2(2\t)+\s_3(2\t)\right]
+X(\t_-)\,\s_1(2\t)=\nn \\
&&\phantom{X(\t_-)\,X(\t_+)}=\left[\s_1(2\t)+\s_2(2\t)+\s_3(2\t) \right] +
 (N-1)\,\s_1(2\t)\,X(\t_+)\, .
\label{bootcc}
\eea
As we have checked, equation \ref{bootcc} turns out to be an identity.
Notice that in this equation all information needed is the ratio $R_1/R_2$,
which is fixed by the BYBE, and the ratio between bulk \SM elements (CDD 
factors cancel out) which is fixed by the bulk Yang-Baxter equation; in other 
words, it depends only on the $O(N)$ structure and not at all on CDD factors. 
This is quite surprising, since the consistency check is meaningful 
only if bound-states exist, which instead 
depends crucially on the presence of the CDD factor.

The explicit expression for $R_B^B(\t)$ is
\be
R_B^B(\t)= -\frac{(i\l+2\t)(i \p -i\l+2\t)}
{(2\t)(i \p -i\l-2\t)} R_1(\t_-)R_1(\t_+)\s_2(2\t)\, .
\ee
Similar bootstrap computations give the reflection amplitudes for the 
antisymmetric tensor components,
\bea
&&R_{B_{ij}}^{B_{ij}}(\t)=
-\frac{(i\l+2\t)}{2\t}R_1(\t_-)R_1(\t_+)\s_2(2\t)\, , \quad i,j\neq N\, ,\\
&&R_{B_{iN}}^{B_{iN}}(\t)=-\frac{(i\l+2\t)(i \p -i\l+2\t)}
{(2\t)(i \p -i\l-2\t)} R_1(\t_-)R_1(\t_+)\s_2(2\t)\, .
\eea
These amplitudes satisfy unitarity automatically. In this case 
boundary crossing-unitarity reads
\bea
&&K^{BB}(\t)=K^{BB}(-\t)S_{BB}^{BB}(2\t)+
\frac{(N-1)(N-2)}{2}K^{B_{ij}B_{ij}}(-\t) S_{B_{ij}B_{ij}}^{BB}(2\t)+ \nn \\
&&\phantom{K^{BB}(\t)=}+(N-1)K^{B_{iN}B_{iN}}(-\t) 
S_{B_{iN}B_{iN}}^{BB}(2\t) \ . \
\label{2bcu} 
\eea
and we have checked that it is indeed satisfied.

\resection{General Boundary Conditions}

For the general boundary condition \ref{bc} let us assume that for
$i=1,\ldots, M$, $\e_i=+1$ and for $i=M+1,\ldots, N$, $\e_i=-1$, which
is, up to index reshuffling, the most general boundary condition we have
to consider.
Since, as before, the boundary action has no term involving flavor
change, we have to assume that the reflection matrix is diagonal. 
Therefore we have the following ansatz for the non-vanishing elements of the 
reflection matrix\footnote{From now on we will use letters in the middle 
of the alphabet $(i,j,...)$ for fermions labeled from $1$ to $M$, and letters 
in the beginning of the alphabet $(a,b,...)$ for fermions labeled from 
$M+1$ to $N$.}
\be
R_i^i(\t)=R_1(\t)\, , \qquad \  \ \qquad R_a^a(\t)=R_2(\t) \ . \
\ee
Similarly as in previous case, the BYBE fixes the ratio $R_1(\t)/R_2(\t)$.
Consider first the factorization of the two-particle reflection process
$|A_a(\t)\,A_i(\t')\rangle_{in} \rightarrow |A_i(-\t')\,A_a(-\t)\rangle_{out}$;
it gives the following equation:
\bea
&&\s_2(\t_{12})\,R_1(\t_1)\,\s_3(\bar\t_{12})\,R_1(\t_2)+
\s_3(\t_{12})\,R_2(\t_1)\,\s_2(\bar\t_{12})\,R_1(\t_2)=  \\
&&\phantom{\s_2(\t_{12})}=
R_2(\t_2)\,R_1(\t_1)\,\s_2(\bar\t_{12})\,R_1(\t_1)\,\s_3(\t_{12})+
R_2(\t_2)\,\s_3(\bar\t_{12})\,R_2(\t_1)\,\s_2(\t_{12})\, , \nn
\eea
where $\bar \t_{12}=\t_1+\t_2$.
By dividing this expression by $R_2(\t_1)\,R_2(\t_2)\,\s_2(\t_{12})\,
\s_2(\bar\t_{12})$ and taking the limit $\t_1 \rightarrow \t_2$ we get a 
differential equation for $X(\t)=R_1(\t)/R_2(\t)$,
\be
\frac{d}{d\t}X(\t)=\frac{X^2(\t)-1}{2\t} \ , \
\ee
whose solutions are
\be
X(\t)=\frac{C-\t}{C+\t}\,  , \qquad \mbox{and} \qquad X(\t)=1 \ , \
\ee
$C$ being an arbitrary integration constant. The solution $X(\t)=1$ corresponds
to Neumann {\em or} Dirichlet boundary conditions, since in these cases 
$R_1(\t)=R_2(\t)$, which we have analyzed in the 
previous section\footnote{Recall 
that the Neumann and Dirichlet cases will only differ by overall CDD 
factors.}. 
 
Consider now the factorization of the process 
$|A_a(\t)\,A_a(\t')\rangle_{in} \rightarrow 
|A_a(-\t')\,A_a(-\t)\rangle_{out}$. It gives the following equation:
\bea
&&\left[M\s_1(\t_{12})+\s_2(\t_{12})+\s_3(\t_{12})\right]R_1(\t_1)\,
\s_1(\bar\t_{12}\, R_2(\t_2) +\nn \\
&&\phantom{R_1(\t_1)\s_1}+ \s_1(\t_{12})\,R_2(\t_1)
\left[(N-M)\s_1(\bar\t_{12})+\s_2(\bar\t_{12})+\s_3(\bar\t_{12})\right]
R_2(\t_2)=\nn\\
&&=R_1(\t_2)\left[M\s_1(\bar\t_{12})+\s_2(\bar\t_{12})+\s_3(\bar\t_{12})\right]
R_1(\t)\,\s_1(\t_{12})+ \nn\\
&&\phantom{R_1(\t_1)\s_1}+R_1(\t_2)\, \s_1(\bar\t_{12})\,R_2(\t_1)
\left[(N-M)\s_1(\t_{12})+\s_2(\t_{12})+\s_3(\t_{12})\right] \, .
\eea
If we plug $X(\t)$ in this expression, the final, compact result is that
$C$ is fixed to be 
\be
C=-\frac{i\pi}{2}\frac{N-2M}{N-2}\, . \label{C}
\ee
By analyzing the BYBE for the other processes we find that \ref{C} 
is the unique
consistent solution. Notice that by taking $M=N-1$ in equation \ref{C}, 
we obtain $C=i\pi/2$ and, for the \nls, we recover Ghoshal's results for 
fixed bc.
All that is left to do is to compute the prefactors
for the reflection amplitudes, by using unitarity and boundary 
crossing-unitarity \ref{RConditions}, 
which will fix $R_1(\t)$ and $R_2(\t)$ up to CDD factors. These conditions
read explicitly
\bea
&&R_1(\t)R_1(-\t)=1 \qquad \ , \ \qquad R_2(\t)R_2(-\t)=1\, , \\
&&R_1(\frac{i\pi}{2}-\t)=
\left[\frac{i\l-2\t}{i\l+2\t}\right]
\left[\frac{i\l(N-M-1)-2\t}{i\l(N-M-1)+2\t}\right]
\sigma_I(2\t)\, R_1(\frac{i\pi}{2}+\t)\, ,
\label{gen}
\eea
where $\s_I(\t)=N\s_1(\t)+\s_2(\t)+\s_3(\t)$. 
The minimal solution of equations
\ref{gen} is given by 
\bea
&&R_1(\t)= - f(\t) R_0(\t)\,
\frac{\Gamma(\frac{1}{4}+\frac{\l}{4\pi}+\frac{i\t}{2\pi})\,
      \Gamma(\frac{3}{4}+\frac{\l}{4\pi}-\frac{i\t}{2\pi})}
     {\Gamma(\frac{1}{4}+\frac{\l}{4\pi}-\frac{i\t}{2\pi})\,
      \Gamma(\frac{3}{4}+\frac{\l}{4\pi}+\frac{i\t}{2\pi})}
\frac{\Gamma(\frac{1}{4}+\frac{\l}{4\pi}(N-M-1)+\frac{i\t}{2\pi})\,}
     {\Gamma(\frac{1}{4}+\frac{\l}{4\pi}(N-M-1)-\frac{i\t}{2\pi})}\times \nn \\
&&\phantom{R_1(\t)=-fR_0(\t)\,}
\times\frac{\Gamma(\frac{3}{4}+\frac{\l}{4\pi}(N-M-1)-\frac{i\t}{2\pi})}
     {\Gamma(\frac{3}{4}+\frac{\l}{4\pi}(N-M-1)+\frac{i\t}{2\pi})}\, ,
\eea
and 
\bea
&&R_2(\t)= - f(\t) R_0(\t)\,
\frac{\Gamma(\frac{1}{4}+\frac{\l}{4\pi}+\frac{i\t}{2\pi})\,
      \Gamma(\frac{3}{4}+\frac{\l}{4\pi}-\frac{i\t}{2\pi})}
     {\Gamma(\frac{1}{4}+\frac{\l}{4\pi}-\frac{i\t}{2\pi})\,
      \Gamma(\frac{3}{4}+\frac{\l}{4\pi}+\frac{i\t}{2\pi})}
\frac{\Gamma(\frac{1}{4}+\frac{\l}{4\pi}(N-M-1)+\frac{i\t}{2\pi})\,}
     {\Gamma(\frac{1}{4}+\frac{\l}{4\pi}(N-M-1)-\frac{i\t}{2\pi})}\times \nn \\
&&\phantom{R_1(\t)=-fR_0(\t)\,}
\times\frac{\Gamma(-\frac{1}{4}+\frac{\l}{4\pi}(N-M-1)-\frac{i\t}{2\pi})}
     {\Gamma(-\frac{1}{4}+\frac{\l}{4\pi}(N-M-1)+\frac{i\t}{2\pi})}\, ,
\eea
where $f(\t)$ is given in equation \ref{fdef2} and $R_0$ in equation 
\ref{Gfc}. The same amplitudes, but with $f\,R_0$ replaced by $R_0$, apply to 
the \nls model.

Notice that $R_2$ has a pole at $\t=-\frac{i\l}{4}(N-2M)$, which is in 
the physical strip for $N/2 < M \leq N-1$. As argued by Ghoshal, this pole 
signals the presence of one-particle contributions in the boundary state.
Since upon the substitution $M\rightarrow N-M$, the 
ratio $X(\t) \rightarrow 1/X(\t)$ and $R_1$ and $R_2$ 
get interchanged, for $1<M<N/2$ it will be $R_1$ to exhibit
this pole in the physical strip.

It is interesting to notice also that for $M=N$, $R_1$ reduces to
the second minimal solution for the free boundary condition, equation 
\ref{fdef3}.

The amplitudes for the two-particle bound-states
can be computed by boundary-bootstrap, and they read explicitly
\bea
&&R_{B}^{B}(\t)=-\frac{(i\l+2\t)(2C -i\l+2\t)}
{(2\t)(2C -i\l-2\t)} R_1(\t_-)R_1(\t_+)\s_2(2\t)\, ,\\
&&R_{B_{ij}}^{B_{ij}}(\t)=
-\frac{(i\l+2\t)}{2\t}R_1(\t_-)R_1(\t_+)\s_2(2\t)\, ,\\
&&R_{B_{ab}}^{B_{ab}}(\t)=-\frac{(i\l+2\t)(2C -i\l+2\t)(2C +i\l+2\t)}
{(2\t)(2C -i\l-2\t)(2C +i\l-2\t)} R_1(\t_-)R_1(\t_+)\s_2(2\t)\, ,\\
&&R_{B_{ia}}^{B_{ia}}(\t)=-\frac{(i\l+2\t)(2C -i\l+2\t)}
{(2\t)(2C -i\l-2\t)} R_1(\t_-)R_1(\t_+)\s_2(2\t)\, ,
\eea
In this more general case, the appropriate bootstrap consistency
condition corresponding to the generalization of equation \ref{bootcc} is
\bea
&&X(\t_-)X(\t_+)[M\s_1(2\t)+\s_2(2\t) +\s_3(2\t) ]+X(\t_-)(N-M)\s_1(2\t)=\nn \\
&&=[(N-M)\s_1(2\t) +\s_2(2\t) +\s_3(2\t)]+M\s_1(2\t)X(\t_+)\, .
\eea
It is very easy to see that if $M=N-1$ this reduces to equation \ref{bootcc}.
As before we should stress that bootstrap consistency requires only information
obtained from the BYBE, via the ratio $X(\t)$, and the fact that the
elementary fermions form isoscalar bound-states.
Finally, we have also explicitly checked that the reflection 
amplitudes listed above satisfy the appropriate boundary crossing-unitarity.

\section{Conclusions}

In this paper we have computed the minimal boundary \SM for the
Gross-Neveu model, extending Ghoshal's analysis of the nonlinear $\s$
model. We found general (diagonal) solutions for the boundary
Yang-Baxter equation for both models and connected them to the
boundary conditions proposed recently in our paper \cite{MorDeMar}. We
also proved that the solutions presented in this paper are consistent
with the boundary-bootstrap. This seems to indicate that the boundary
contains a lot of information of the bulk theory. It would be
interesting to investigate how much we can learn about bulk integrable
models {\em starting} from the boundary.  As a natural follow-up to
this work it would be interesting to study the boundary Yang-Baxter
equation in general and see if there are non-diagonal solutions, that
is flavor-changing scattering off the boundary. There should be
possible, then to find associated microscopic boundary conditions for
the elementary fields.

\section*{Acknowledgments}
We would like to thank D.~Jatkar and S.F.~Hassan for useful discussions. 

\newpage
\appendix
\resection{Appendix}

In this appendix we list the \SM elements for scattering processes
involving two-particle bound-states 
(the isoscalar and the first antisymmetric tensor), 
that we need to write down the boundary crossing-unitarity equation
\ref{1bcu}, \ref{2bcu}. These amplitudes are obtained from 
those of elementary fermions by fusion. 

Recall that if the scattering amplitude of two particles $A_a$ and $A_b$
has a simple pole in the physical strip at, say, $\t=iu_{ab}^c$, 
with residue given by :
\be
S_{ab}^{a'b'}(\t)\simeq \frac{if_{ab}^c f_c^{a'b'}}{\t-iu_{ab}^c}\, ,\nn
\ee
then the scattering of the bound-state particle $A_c$ with all other 
particles in the theory can be obtained by the bootstrap equation \cite{ZZAnn}
\be
f_{ab}^cS_{cd}^{c'd'}(\t)=f_{a_1b_1}^{c'}S_{bd}^{b_1d_1}
(\t-i\bar u_{b\bar c}^{\bar a})\, S_{ad_1}^{a_1d'}
(\t+i\bar u_{\bar c a}^{\bar b})\, ,\nn
\ee
where $\bar u\equiv \pi-u$. 
In the case of GN model, we get
\bea
S_{BB}^{BB}(\t)&=&\frac{\t(i\pi-\t)\left[\l(\pi-3\l)+\t(i\pi-\t)\right]
-2\pi\l^2(\l-\pi))}{\t(i\pi-\t)(\t-i\l)(i\pi-\t -i\l)}
\s_2(\tilde\t_+)\,\s_2(\tilde\t_-)\,\s_2^2(\t) \, , \nn \\
S_{BB}^{B_{ij}B_{ij}}(\t)&=&\frac{2i(N-4)\l^3}
{(i \p-\t)(\t-i\l)(i \p-\t-i\l)} \s_2(\tilde\t_+)\,\s_2(\tilde\t_-)\,
\s_2^2(\t) \, ,\nn \\
S_{B_{ij}B_{ij}}^{BB}(\t)&=&\frac{i(N-4)\l^3}
{2(i \p-\t)(\t-i\l)(i \p-\t-i\l)} \s_2(\tilde\t_+)\,\s_2(\tilde\t_-)\,
\s_2^2(\t) \, , \nn \\
S_{BB_{ij}}^{BB_{ij}}(\t)&=&\frac{\t(i \p-\t)+\l(\pi-3\l)}
{(\t-i\l)(i \p-\t -i\l)} \s_2(\tilde\t_+)\,\s_2(\tilde\t_-)\,\s_2^2(\t) \, , 
\nn \\
S_{BB_{ij}}^{B_{ij}B}(\t)&=&\frac{-i(N-4)\l^3}
{\t(\t-i\l)(i \p-\t-i\l)} \s_2(\tilde\t_+)\,\s_2(\tilde\t_-)\,\s_2^2(\t) \,
\ , \ 
\nn 
\eea
where $\tilde\t_\pm=\t\pm i\l$. Unitarity follows directly from the fusion 
procedure and crossing symmetry is satisfied with charge conjugation matrix
elements $C^{BB}=1$ and $C^{B_{ij}B_{ij}}=-2$.

\end{document}